\documentclass[showpacs,twocolumn,aps,prb,floatfix,superscriptaddress,10pt]{revtex4-1}
\usepackage{graphicx}
\usepackage{color}

\begin{document}
\title{Quantum transport in chemically functionalized graphene at high magnetic field:\\ 
Defect-Induced Critical States and Breakdown of Electron-Hole Symmetry}

\author{Nicolas Leconte}
\affiliation{ICN2 - Institut Catala de Nanociencia i Nanotecnologia, Campus UAB, 08193 Bellaterra (Barcelona), Spain}
\affiliation{Universit\'{e} catholique de Louvain (UCL), Institute of Condensed Matter and Nanosciences (IMCN), Chemin des \'{e}toiles 8, 1348 Louvain-la-Neuve, Belgium}
\email{nicolas.leconte@icn.cat}
\author{Frank Ortmann}
\affiliation{ICN2 - Institut Catala de Nanociencia i Nanotecnologia, Campus UAB, 08193 Bellaterra (Barcelona), Spain}
\author{Alessandro Cresti}
\affiliation{IMEP-LAHC (UMR CNRS/INPG/UJF 5130), Grenoble INP, Minatec, 3, Parvis Louis N\'eel, CS 50257, F-38016, Grenoble, France}
\author{Jean-Christophe Charlier}
\affiliation{Universit\'{e} catholique de Louvain (UCL), Institute of Condensed Matter and Nanosciences (IMCN), Chemin des \'{e}toiles 8, 1348 Louvain-la-Neuve, Belgium}
\author{Stephan Roche}
\affiliation{ICN2 - Institut Catala de Nanociencia i Nanotecnologia, Campus UAB, 08193 Bellaterra (Barcelona), Spain}
\affiliation{ICREA - Institucio Catalana de Recerca i Estudis Avan\c cats, 08010 Barcelona, Spain}
\date{\today}

\begin{abstract}
  Unconventional magneto-transport fingerprints in the quantum Hall regime (with applied magnetic field from one to several tens of Tesla) in chemically functionalized graphene are reported. Upon chemical adsorption of monoatomic oxygen (from $0.5 \%$ to few percents), the electron-hole symmetry of Landau levels is broken, while a double-peaked conductivity develops at low-energy, resulting from the formation of critical states conveyed by the random network of defects-induced impurity states. Scaling analysis hints towards the existence of an additional zero-energy quantized Hall conductance plateau, which is here not connected to degeneracy lifting of Landau levels by sublattice symmetry breakage. This singularly contrasts with usual interpretation, and unveils a new playground for tailoring the fundamental characteristics of the quantum Hall effect. 
\end{abstract}

\maketitle


\section{Introduction}

In graphene, the electron-hole symmetry and the coexistence of electrons and holes at the Dirac point play an important role in the concept of massless Dirac fermions \cite{CastroNeto:2009cl}, which is intimately connected to a large class of intriguing quantum transport phenomena such as Klein tunneling \cite{Katsnelson:2006p5778}, weak-antilocalization \cite{McCann:2006ip,Tikhonenko:2009ho} and anomalous quantized Hall conductivity \cite{Novoselov:2005es,Zhang:2005gp,Goerbig:2011bh}. In the high magnetic field regime, graphene exhibits a symmetric electron/hole energy spectrum, which includes a zero energy Landau level (where electrons and holes coexist), and non-equidistant Landau levels (LLs), with energies $\varepsilon_{n}=sgn(n)\sqrt{2\hbar{v_{F}}^{2}eB|n|}$, where $v_{F}=10^{6}{\rm m}{\rm s}^{-1}$ is the Fermi velocity, $B$ is the magnetic field and $n$ is the integer LL index~\cite{McClure:1956vp,CastroNeto:2009cl}. 

Several symmetry-breaking mechanisms can lift the four-fold degeneracy of LLs \cite{Rickhaus:2012br,Barlas:2012et}. In particular, the presence of an additional quantized Hall plateau $\sigma_{xy}=0$ at low energy in high-mobility samples has been tentatively analyzed in terms of energy-splitting (gap) of the zero-energy LL, either driven by Zeeman interaction (spin degeneracy lifting) or the formation of quantum Hall ferromagnetism  \cite{Zhang:2006hn,Nomura:2006bi,Young:2012bn}. The scaling of the zero-energy split gap with magnetic field remains however theoretically and experimentally debated, with a behavior changing from linear to $\sqrt{B}$-like, depending on the underlying dominant symmetry breaking or on the sample quality \cite{Jung:2009hm,Young:2012bn}. In highly disordered graphene, a crossover from the Efros-Shklovskii to Mott variable-range hopping has been reported and related to a localization length exceeding the screening length, thus elucidating the nature of transport at the Hall plateau transition and emphasizing the important role of disorder \cite{Bennaceur:2012eu}.  The existence of a metal-insulator transition (MIT) is actually a central aspect of the quantum Hall effect (QHE) theory, which requires the robustness of electronic states at the center of Landau level, which have peculiar multifractal properties ({\it critical states}) \cite{Aoki:1987do,OST_PRB77}. The recent observation of a quantized Hall conductance in highly resistive (millimeter-scale) hydrogenated graphene, with mobility less than $10 {\rm cm}^{2}/{\rm V.s}$ and estimated mean free path far beyond the Ioffe-Regel limit, even suggests some unprecedented robustness of the QHE in strongly damaged graphene, a fact that crucially demands for theoretical inspection and analysis of the related MIT \cite{GUI_PRL110}.

On the other hand, in absence of magnetic fields, adsorbed impurities and defects are found to produce a large variety of additional transport features in graphene, such as resonant tunneling, mobility gaps and electron-hole asymmetry (see for instance \cite{Cresti:2008il,Wehling:2009wb,Leconte:2011hs,Roche:2012ji}). Available high-field magnetotransport experiments in graphene also frequently present some weak electron-hole asymmetry together with some conductivity variability~\cite{Giesbers:2009if,Kurganova:2011bc,Zhao:2012eh}, which could hint towards a signature of specific impurities. However, no connection with specific disorder has been established so far. Besides, the possibility of tuning QHE fingerprints by intentional disorder (chemical functionalization, controlled doping,...)\cite{Boukhvalov:2008ix,Lu:2009gc,Sun:2011ks} would certainly allow a richer exploration of the metal-insulator transition and the QHE phase diagram in two-dimensional systems, which could be interesting for metrology issues~\cite{Weiss:2008er,Janssen:2011gt,Guignard:2012cy}.

The present Letter reports on high-field magnetoconductivity fingerprints of graphene functionalized by oxygen (epoxy) defects, using highly efficient quantum transport simulations. The complex interplay between chemical disorder and magnetic field on the MIT and the QHE are studied for realistic magnetic field strengths (from a few to several tens of Tesla). For an epoxy defect concentration as low as $0.5\%$, a strong electron-hole magnetotransport asymmetry is observed, with an almost complete suppression of Landau levels for electrons, in contrast to the robustness of quantum Hall regime for hole conduction. Most importantly, a low concentration of epoxy defects induces magnetoconductivity split peaks around the zero energy, whose separation scales almost linearly with the magnetic field. This observation, together with the analysis of the energy-dependent conductivity scaling behavior, indicates the formation of critical states and mobility edges. These remarkable high-field magnetotransport characteristics should be accessible by experiments using soft chemical treatment of graphene such as atomic oxygen deposition in ultrahigh vacuum \cite{Hossain:2012cq} or ozone exposure \cite{Moser:2010us}, which preferentially produce epoxy defects. 

\section{Methods} 

To explore quantum transport in oxygen-functionalized graphene, first a simple $\pi$-$\pi$* orthogonal tight-binding (TB) model with nearest neighbors hopping of $2.6$ eV is used for pristine graphene, represented by the Hamiltonian $H_0$. Following a previous {\it ab initio} study based on a supercell approach \cite{Leconte:2011hs, Leconte:2010p4412}, two additional orbitals (p$_{\rm x}$ and p$_{\rm z}$) {\it per defect} are included to account for the presence of oxygen adatoms in C--C bridge position. Electronic parameters between oxygen atoms and their nearest neighbor carbon atoms (denoted $p$ and $q$) are introduced as
\begin{eqnarray}
H^{\rm epoxy}_{pq}&=&\epsilon_1 \left(c_p^\dag c_p + c_q^\dag c_q \right) + \epsilon_x d_x^\dag d_x + \epsilon_z d_z^\dag d_z\\
&&+\left[ \gamma_x \left(c_p^\dag d_x - c_q^\dag d_x \right) +\gamma_z\left(c_p^\dag d_z + c_q^\dag d_z \right)+ {\rm h.c.} \right]  \nonumber
\end{eqnarray}
for zero magnetic field \cite{Leconte:2010p4412}, where $d_x$ and $d_y$ are the annihilation operators for the extra orbitals located in bridge position, $\epsilon_1=1.5$ eV, $\epsilon_x=-2.5$ eV, $\epsilon_z=-1$ eV, $\gamma_x=-4.68$ eV, $\gamma_z=3.9$ eV, yielding the total Hamiltonian $H=H_0 +\sum_{\text{epoxy}} H^{\text{epoxy}}$. The effect of the external magnetic field is taken into account through a standard Peierls phase factor on the hopping elements of the Hamiltonian~\cite{Luttinger:1951ua, Zhu:2011bz, Yuan:2012dg}. High-magnetic field transport is studied with an order-$N$, real space implementation of the Kubo approach for $\sigma(E,B)$ \cite{Ortmann:2011fz, OrtmannArxiv}. The scaling properties of $\sigma$ can be followed through the dynamics of electronic wavepackets using $\sigma(E,t)=e^{2}\rho(E)\Delta X^{2}(E,t)/t$, where $\rho(E)$ is the density of states, $\Delta X^{2}(E,t) = Tr\left[\delta(E-\hat{H})\left|\hat{X}(t)-\hat{X}(0)\right|^2\right]/Tr\left[\delta(E-\hat{H})\right]$ (with $\hat{X}(t)$ the position operator in Heisenberg representation) is the energy- and time-dependent mean quadratic displacement of the wave packet.
Calculations are performed on systems containing more than ten million carbon atoms, which corresponds to sizes larger than $500\times 500~{\rm nm}^{2}$, and randomly distributed defects.
In all simulations, the energy smearing factor is $1.5$ meV. 

\section{Results}

\subsection{Disorder-induced QHE asymmetry}

\begin{figure}[tb]
  \resizebox{8cm}{!}{\includegraphics{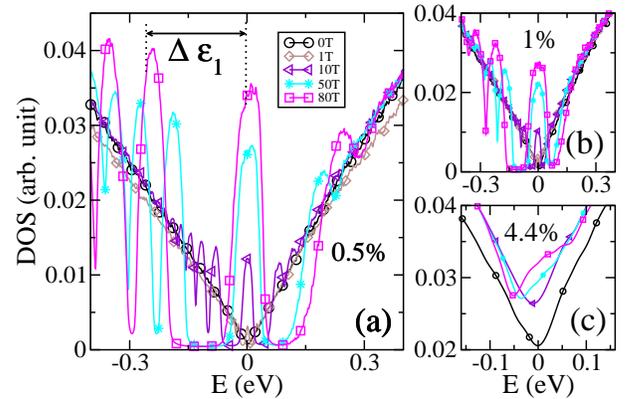}}
  \caption{(color online) Density of states for various magnetic fields $B$ and oxygen concentrations. Epoxy defect concentrations are 0.5$\%$ (a), 1$\%$ (b) and 4.4$\%$ (c) whereas the magnetic field ranges from 1 Tesla to 80 Tesla.}
  \label{fig1}
\end{figure}

The evolution of the electronic structure under high magnetic fields is first scrutinized by varying the density of adsorbed oxygen impurities. Figure~\ref{fig1} presents the density of states (DOS) for three different epoxy defect densities, namely $0.5 \%$ (a), $1 \%$ (b) and $4.4 \%$ (c).
For the lowest concentration of $0.5\%$, magnetic fields higher than $1$ T are required to observe the onset of the zero-energy LL, whereas no LL signature is observed at higher energies at $1$ T. By further increasing the field to 10, 50 and 80 T, the sequence of LLs clearly develops on the hole side of the spectrum, with distances between LLs ($\Delta\epsilon_{n}$) in full agreement with the expected $\sqrt{B}$ spacing (for example $\Delta\epsilon_1 \approx 0.3$ eV at 80 T)\cite{McClure:1956vp}. In contrast, LLs only partially develop in the electron region of the spectrum. The $1\%$ density is also shown to inhibit any LL on the electron side of the spectrum, whereas the formation of LLs in the hole side becomes more difficult and requires higher magnetic fields. For densities as high as $4.4 \%$, a total disappearance of LLs is observed throughout the whole spectrum and even fields as large as $80$ T fail to generate the zero-energy level.

\subsection{Magnetoconductivity fingerprints}

\begin{figure}[tb]
  \resizebox{8cm}{!}{\includegraphics{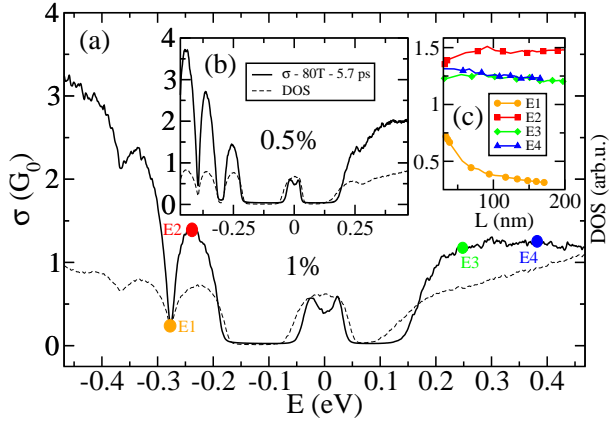}}
  \caption{(color online) (a) $\sigma(E)$ in units of the conductance quantum for $1\%$ epoxy defect density. (b) $\sigma(E)$ for $0.5\%$ epoxy defect density. Dashed lines denote the corresponding (rescaled) DOS in both panels. (c) Length dependent $\sigma(L)$ at four selected energies indicated in (a).}
 \label{fig2}
\end{figure}

The conductivity $\sigma$ is further analyzed for several representative defect densities and magnetic fields. Figures~\ref{fig2} (a) and (b) show $\sigma(E)$ in units of the conductance quantum ($G_{0}=2e^{2}/h$), computed at the largest accessible time scale ($t=5.7$ ps) for impurity densities of $1 \%$ (a) and $0.5 \%$ (b) at $B=80$T. The rescaled \textit{B}-dependent DOS is superimposed (dashed lines) for the sake of comparison. 

The electron-hole asymmetry reported for the DOS consistently develops for the conductivity as well, and is even more pronounced. 
It is worth noticing that $\sigma(E)$ downscales with the defect density for sufficiently high electron energies  (compare for instance Figs.~\ref{fig2} (a) and (b) at $E=0.4$ eV), whereas it hardly changes for the lowest energy LL0, as well as for the first (LL-1) and second (LL-2) Landau levels on the hole side of the spectrum.

Figure~\ref{fig2}(c) provides a clearer insight into the transport properties on the electron and hole side of the spectrum by showing the length-dependent conductivity $\sigma$ at $80$T for the selected electron energies $E_1$ to $E_4$ indicated in Fig.~\ref{fig2}(a).
On the electron side, at $E_3$ and $E_4$, the conductivity slowly decreases with $L$, thus indicating the transition from diffusive to localized regime. This is consistent with the impeded formation of LLs in this energy region. On the hole side of the spectrum, $E_2$ corresponds to the $\sigma$ peak (LL-1 center) and $E_1$ corresponds to the $\sigma$ minimum in between LL-1 and LL-2. At $E_2$ the conductivity does not decrease with $L$, thus clearly confirming that the states remain delocalized. On the contrary, at $E_1$ states localize with a rapidly decreasing $\sigma$. 

\begin{figure}[tb]
  \resizebox{8cm}{!}{\includegraphics{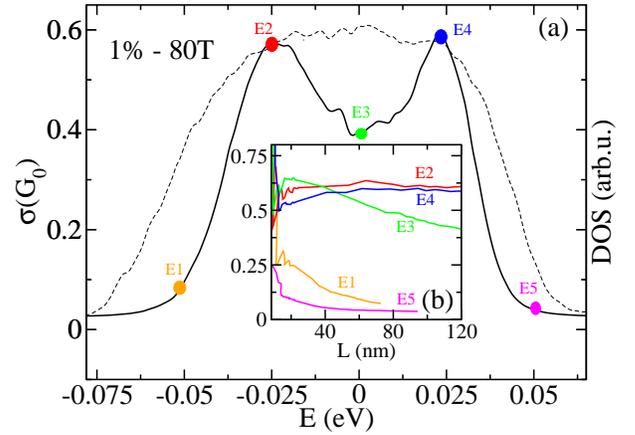}}
  \caption{(color online) 
	(a) Structure of the double peaked $\sigma(E)$ in units of the conductance quantum for $1\%$ epoxy defect density (solid lines). Superimposed rescaled DOS in dotted lines.
	(b) Length dependent $\sigma(L)$ at the five selected energies indicated by colored points in (a).}
  \label{fig3a}
\end{figure}

\begin{figure}[tb]
  \resizebox{8cm}{!}{\includegraphics{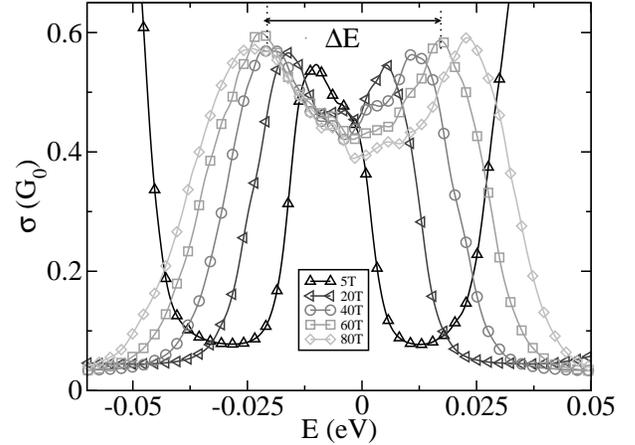}}
  \caption{$\sigma(E)$ for $1\%$ epoxy defects and varying the field.}
  \label{fig3b}
\end{figure}

The main finding of the study is described in Fig.~\ref{fig3a} and Fig.~\ref{fig3b}.  The peculiar feature [already observed in Figs.~\ref{fig2} (a) and (b)] is the doubly peaked conductivity seen in the lowest energy LL0, whose degeneracy is however not lifted by (defect-induced) sublattice symmetry breakage.  Actually, the conductivity scaling in Fig.~\ref{fig3a}(b) at the two peaks ($E_2$ and $E_4$) from Fig.~\ref{fig3a}(a) remains length-independent, whereas the decay of $\sigma$ at energies $E_1$, $E_3$ and $E_5$ (where the DOS remains finite) clearly evidences the localized nature of corresponding states and formation of insulating regime. Such a feature indicates the existence of some MIT and the formation of critical states in between which a zero-energy Hall conductivity plateau possibly could develop~\cite{OST_PRB77,Evers:2008gi}. This is entirely different from the LL-1 discussed above. Additional transverse conductivity calculations are required to confirm this new zero energy plateau, but are out of the scope of the present study.

Figure~\ref{fig3b} gives the further evolution of the double-peaked conductivity $\sigma(E)$ with increasing magnetic field for 1$\%$ epoxy defect concentration. Figure~\ref{fig3c} shows that the magnetic field dependence of the peak separation $\Delta E(B) = E_4(B) - E_2(B)$ exhibits a steeper slope for higher epoxy density ($1.77\%$). The different trends between lower and higher impurity concentrations thus offer an experimental test when measuring QHE in weakly oxidized and disordered graphene. Possible sublinear magnetic field dependence might serve as an additional hint. Proper characterization of the samples might also rule out between this and other explanations from the literature. However, such a surprising behavior cannot be explained in terms of sublattice-symmetry breaking arguments and demands for an alternative interpretation. The value of $\Delta E$ at zero field has a different, more conventional, origin. Indeed, it represents the separation between conventional diffusive (or weakly localized) states by states (E=0) with stronger localization behavior, which is observed for moderate concentration of nonresonant adsorbates~\cite{PhysRevLett.111.146601}. This conductivity dip was not observed in our previous zero-field study~\cite{Leconte:2011hs} due to the larger numerical smearing of states.

\subsection{Origin of the double-peak conductivity structure}  

\begin{figure}[tb]
  \resizebox{8cm}{!}{\includegraphics{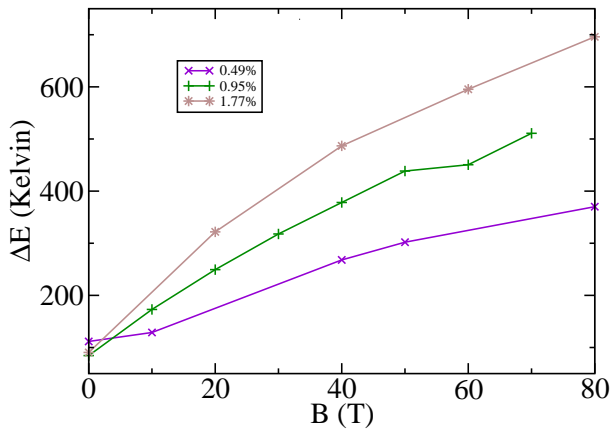}}
  \caption{(color online) 
  Peak separation $\Delta E(B)=E_{4}(B)-E_{2}(B)$ for different defect concentrations.}
  \label{fig3c}
\end{figure}

\begin{figure}[tb]
  \resizebox{8cm}{!}{\includegraphics{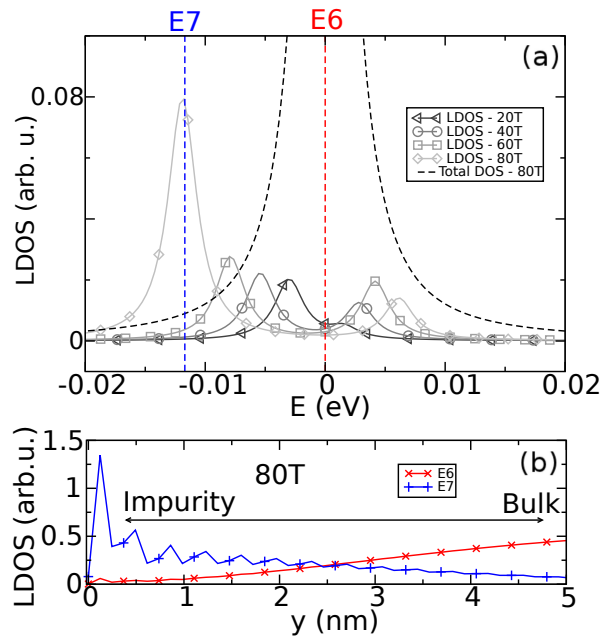}}
  \caption{(color online) (a) A single oxygen atom on pristine graphene sheet: total DOS and LDOS for a bridging carbon atom at different magnetic fields. (b) Spatial dependency of LDOS by distance from the oxygen atom (at $y=0$) at $80$T.}
  \label{fig4a}
\end{figure}

Under magnetic field, the electronic structure of pristine graphene consists of well-separated degenerate flat LLs. The DOS of the lowest LL is peaked at zero energy, while the corresponding local DOS (LDOS) is spatially uniform over the system. When a single epoxy impurity is introduced, two peaks (one at negative energy and a smaller one at positive energy) appear in the LDOS close to the oxygen atom (at one of the bridging carbon atoms), as shown in Fig.~\ref{fig4a} for different magnetic fields. This indicates the formation of two impurity-bound states. The energy separation between these two states in Fig.~\ref{fig4a}(a) increases roughly linearly with $B$. We recall that we observed a similar linear increase for the two \textit{conductivity} peaks of Fig.~\ref{fig3b}. The similar magnetic field scaling is a first indication of the importance of these single-impurity-pinned states, which are investigated in more detail below.

First, the locality of such states is analyzed. In Fig.~\ref{fig4a}(b), the LDOS is plotted for the carbon atoms along a line passing through one of the bridging carbon atoms at $B$=80 T for the two energies $E_6$ and $E_7$ indicated in Fig.~\ref{fig4a}. At $E_6=0$, in between the two peaks, the LDOS corresponding to LL0 turns out to be suppressed over a region with a radius of about $5$ nm around the defect.
At the energy of the peak $E_7$ and in the same spatial region, the LDOS is high and decays rapidly with distance from the epoxy. This observation confirms the bound nature of the corresponding state. Additionally, the spatial extension of the bound state is driven by the magnetic field intensity and it is actually of the same order as the magnetic length ($ l_B \sim 25.7/\sqrt{B}$ nm). 


The connection between the quasi-localized nature of bound states as seen in the LDOS in Fig.~\ref{fig4a}(b) with the formation of extended (critical) states conveying length-independent conductivity is not straightforward and even counterintuitive. We rationalize this intriguing point as follows. These impurity bound states couple to each other, even at relatively low defect density (magnetic lengths are larger than the average distance between impurities). In fact, the overlap between different bound states gives rise to new states, which occur over a certain energy window [the broad DOS feature of LL0 in Fig.~\ref{fig2}(a)]. The width of this window depends on the coupling strength, which increases for higher oxygen concentration and is at the origin of the concentration dependence of the peak separation in Fig.~\ref{fig3c}. Together with the increase of the splitting of states for higher magnetic fields (as seen for the isolated epoxy in Fig.~\ref{fig4a}), this explains the trends observed in Fig.~\ref{fig3c}. Finally the formation of a double conductivity peak in absence of  any splitting of the original LL0 (Fig.~\ref{fig3a}) is thus found not to be related to any symmetry breaking mechanism, but to the occurrence of resonant energies at which percolation takes place between impurity-states, whereas the zero-energy states  are localized inside the fragmented pristine regions, carrying no current, in total contrast to the case of pristine graphene with homogeneous disorder~\cite{Sheng:2006te}.

\subsection{High disorder concentration regime}

\begin{figure}[tb]
  \resizebox{8cm}{!}{\includegraphics{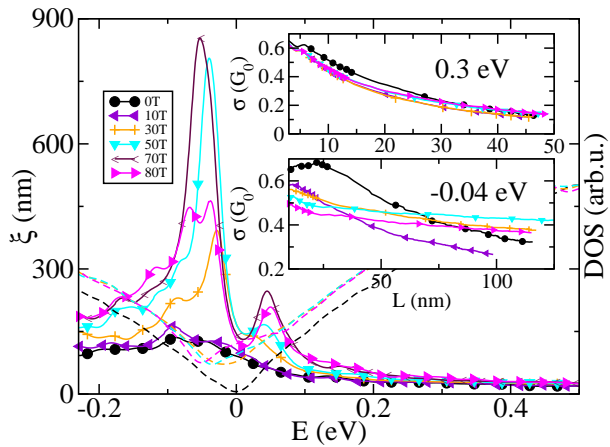}}
  \caption{(color online) Main frame: estimated localization lengths (solid lines) at $4.4\%$ with superimposed rescaled DOS for chosen fields (dashed lines). Inset: length-dependent conductivity scalings at chosen energies.}
  \label{fig2b}
\end{figure}

Finally, the strong disorder limit for epoxy disorder is explored. In Fig.~\ref{fig2b} the localization length $\xi$ is estimated for $4.4\%$ of disorder by fitting $\sigma(L)$ with an exponential law~\cite{Leconte:2011hs}.  The scaling behaviors of $\sigma(L)$ are shown in the inset. The DOS is superimposed for $0$, $30$, $50$ and $80$ Tesla for reference. Surprisingly, for this high disorder case where the formation of a clear LL spectrum is jeopardized in the DOS [see also Fig.~\ref{fig1}(c)], states are not equally (de)localized over the energy spectrum. At zero energy, localized states occur between two bands of more extended states. This is consistent with the previous analysis based on percolation of local impurity states. High magnetic fields are necessary to compensate the conventional disorder induced localization effects (lower inset), which is confirmed by the smaller localization lengths on the electronic side where disorder is stronger (main frame). Similarly, a high-$B$ value is required to resolve the peaks of delocalized states in the spectrum. The localized states seem to be less sensitive to the effect of the magnetic field. On the other hand the field tunes the localization length by up to one order of magnitude in the low energy region (about $40$ meV away from the zero-energy reference). The robustness of such delocalized states at high disorder might explain recent experimental works observing QHE features in highly hydrogenated samples\cite{GUI_PRL110}.

\section{Discussion and Conclusion}

Weak densities of oxygen ad-atoms induce specific QHE features, such as the breakdown of electron-hole symmetry of the high energy LLs. Additionally, despite the absence of a true energy gap, two low-energy conductivity peaks, strongly robust to disorder, are related to the coupling of local impurity-pinned bound states, whose spatial extension and overlap is monitored by the magnetic-field strength. These results suggest the possibility to measure an additional quantized Hall conductance plateau (at $\sigma_{xy}=0$) due to such impurity-pinned local Landau levels.

The specific role of disorder played in this study is different from literature predicting the QHE to be robust under three kinds of disorder, namely (i) weak potentials $V_W$ (satisfying $V_W <  \hbar \omega_c$) \cite{Thouless1981, Halperin1982}, (ii) scattering centers defined by potential $V_{\text{SC}}$ (satisfying a distance $d$ between centers larger than the magnetic length $l_B$) \cite{Prange1982, Joynt1984, Chalker1984} and (iii) smooth potentials $V_S$ (satisfying $\Delta V_S = \frac{\hbar \omega_c}{l_B}$) \cite{Iordansky1982, Floser2013, Prange1982b, Kazarinov1982, Luryi1983, Trugman1983, Giuliani1983, Joynt1984, Tsukada1976}.
 
Epoxy is a scattering center, in the sense that it induces intervalley mixing responsible for quantum localization effects~\cite{Leconte:2011hs}. However, the disorder concentrations in this study do not satisfy the $d \gg l_B$ criterion from (ii). The novelty of this study is that it hints towards the existence of extended states based on a percolation model~\cite{Stauffer1979, Essam1980} in the case of scattering centers, a framework up to now (to the best of our knowledge) only related to smooth potentials~\cite{Iordansky1982, Floser2013, Prange1982b, Kazarinov1982, Luryi1983, Trugman1983, Giuliani1983, Joynt1984, Tsukada1976} where it was also confirmed experimentally~\cite{Floser2013}. Differently from smooth potentials where the critical energy $E_c$ (energy at which the extended state is found) is predicted to be at zero energy for electron-hole symmetric potentials , for epoxy disorder, these critical energies $E^{\pm}_c$ are found at the left and right of $E=0$ eV. The states at energies away from $E^{\pm}_c$ are localized as expected from the literature, possibly leading in this case to a zero energy Hall pleateau. Actually, a similar zero energy Hall plateau has recently been observed in disordered graphene~\cite{{Nam:2013ey}}, but without any ultimate conclusion about its origin. We believe the present work might represent a possible key of interpretation and provide the necessary cornerstone for a better understanding of impurity related critical states. Future calculations might try to generalize the existence of these critical states to a larger class of impurities.

The recent demonstration of chemical control of deposition of monoatomic oxygen atoms on graphene \cite{Hossain:2012cq}, together with several QHE experiments in oxidized and hydrogenated graphene \cite{Pallecchi:2012cs,Pallecchi2013} pave the avenue for in-depth exploration of the QHE in chemically-modified graphene based materials. A low-temperature scanning gate microscopy technique~\cite{Connolly:2012jt} to observe percolation of states in the bulk could shed further light on the specific behavior at these new critical energies.

\section{Acknowledgements.} J.-C.C. and N.L. acknowledge financial support from the F.R.S.-FNRS of Belgium. This research is directly connected to the ARC on 'Graphene Nano-electromechanics' (Nr 11/16-037) sponsored by the Communaut\'e Fran\c caise de Belgique, and funded by the Spanish Ministry of Economy and Competitiveness for funding (MAT2012-33911). Computational resources were provided by the CISM of the UCL. S.R. and F.O. acknowledge the computer resources (from Altamira) provided by the Barcelona Supercomputing Center and the Spanish Supercomputing Network.

\bibliography{file}

\end{document}